\documentclass[twoside,twocolumn]{article}
\pdfoutput=1

\usepackage{graphicx}
\usepackage{amsmath}

\usepackage[hmarginratio=1:1,top=32mm,columnsep=20pt]{geometry}
\usepackage{titlesec}

\usepackage{cite}

\usepackage{bm}      
\usepackage{amssymb} 
\usepackage{authblk}

\newcommand{\fref}[1]{Fig.~\ref{#1}}
\newcommand{\Fref}[1]{Figure~\ref{#1}}

\newcommand{\sref}[1]{Sec.~\ref{#1}}

\newcommand{\rref}[1]{Ref.~\cite{#1}}

\newcommand{\tref}[1]{Table~\ref{#1}}

\renewcommand{\vec}[1]{\bm{#1}} 

\newcommand{\chainlength}{\ensuremath{N}}   

\newcommand{\LJeps}{\ensuremath{\varepsilon}}   
\newcommand{\LJsigma}{\ensuremath{\sigma}}      

\newcommand{\Ns}{\ensuremath{N_{\text{s}}}}       

\newcommand{\coolrate}{\ensuremath{q_{\text{c}}}} 
\newcommand{\Ts}{\ensuremath{T_{\text{s}}}}       
\newcommand{\Ton}{\ensuremath{T_{\text{on}}}}     
\newcommand{\Tg}{\ensuremath{T_{\text{g}}}}       

\newcommand{\isf}{\ensuremath{\phi_{s}^{q}}} 
\newcommand{\tage}{\ensuremath{t_{\text{a}}}} 

\begin{document}

\title{\vspace{-15mm}\fontsize{22pt}{10pt}\selectfont\textbf{%
Inherent Structure Energy is a Good Indicator of Molecular Mobility in Glasses}}

\author[1]{\large Julian Helfferich}
\author[1]{\large Ivan Lyubimov}
\author[1]{\large Daniel Reid}
\author[1]{\large Juan J.\ de Pablo}
\affil[1]{\normalsize %
Institute for Molecular Engineering, %
University of Chicago, %
5640 South Ellis Avenue, %
Chicago, IL 60637, USA. %
E-mail: jhelfferich@uchicago.edu }
\date{}

\twocolumn[%
    \begin{@twocolumnfalse}
\maketitle
\begin{abstract}%
    Glasses produced via physical vapor deposition can display greater kinetic
    stability and lower
    enthalpy than glasses prepared by liquid cooling. While the reduced enthalpy
    has often been used as a
    measure of the stability, it is not obvious whether dynamic measures of
    stability provide the same view. Here, we study dynamics in vapor-deposited and liquid-cooled glass films using molecular simulations
    of a bead-spring polymer model as well as a Lennard-Jones binary mixture in
    two and three
    dimensions. We confirm that the dynamics in vapor-deposited glasses is
    indeed slower than in ordinary glasses. We further show that the
    inherent structure energy is a good reporter of local dynamics,
    and that aged systems and glasses prepared by cooling at progressively slower
    rates exhibit the same behaviour as vapor-deposited materials when they both have the same inherent structure energy. These findings suggest that the
    stability inferred from measurements of the energy
    is also manifested in dynamic observables, and they strengthen the view that vapor
    deposition processes provide an effective strategy for creation of
    stable glasses.

    \bigskip
\end{abstract}%
    \end{@twocolumnfalse}
]

\section{Introduction}
\label{sec:intro}

Ordinary glasses are typically prepared by cooling a liquid at a rate that is
sufficiently fast to avoid crystallization. Upon cooling
towards the glass transition temperature $\Tg$, the viscosity and characteristic
relaxation times
increase considerably~\cite{BerthierBiroli:RMP2011,EdigerHarrowell:JCP2012,%
BiroliGarrahan:JCP2013,Donth:Book2001,BinderKob:Book2011}. Eventually, such
relaxation times exceed the cooling rate, leading to dynamic arrest and glass
formation~\cite{BiroliGarrahan:JCP2013,Donth:Book2001,BinderKob:Book2011}.
This transition is accompanied by a change in the specific heat and defines the
calorimetric glass transition temperature~\cite{EdigerHarrowell:JCP2012,%
Donth:Book2001}. As the relaxation times become larger than the available
laboratory time
scales, the system is no longer able to reach its equilibrium state and thus
``falls out of equilibrium''~\cite{BinderKob:Book2011}. However, dynamics do not
come to a halt in the glass phase. Instead, the system slowly evolves towards
its equilibrium state in a process called ``physical aging'', characterized by
an increase of the density and the structural relaxation time~%
\cite{Struik:Book1978}. Thus, within the traditional view of glass formation,
two strategies can be followed to
prepare a glass that is closer to its equilibrium state: employing a slower
cooling rate, or letting the system age for an extended period period of time.

Recent experiments have shown that glasses can also be created through a process
of physical vapor deposition (PVD), leading to materials whose macroscopic
characteristics, such as mechanical properties or onset temperature, can exceed those of highly aged ordinary glasses.
Specifically, PVD glasses can exhibit extraordinary thermodynamic and kinetic
stability
\cite{SwallenEtal:Science2007,IshiiNakayama:PCCP2014}, an increased density~%
\cite{SwallenEtal:Science2007,%
DalalEdiger:JPCL2012,IshiiNakayama:PCCP2014} and a reduced enthalpy~%
\cite{KearnsEtal:JPCB2008,KearnsEtal:JPCB2009,Leon-GutierrezEtal:TA2009,%
DawsonEtal:JPCL2011}.
A growing body of numerical simulations has sought to interpret from a molecular
perspective available experimental observations on PVD glasses. Simulations have
been able to reproduce several experimentally observed features, including
higher thermodynamic and kinetic stability, the existence of an optimal
substrate temperature for vapor deposition, the existence of a mobile layer at
the vacuum interface, and the ability to control anisotropy through the
deposition process
experiments~\cite{SinghDePablo:JCP2011,SinghEtal:NM2013,LyubimovEtal:JCP2013,%
LinEtal:JCP2014,LyubimovEtal:JCP2015}.

In both experimental and simulation studies, the enthalpy has been used as a
convenient, easily accessible measure for the stability. In experiments, the
enthalpy is
typically determined by relying on calorimetry measurements~\cite{%
KearnsEtal:JCP2007,%
KearnsEtal:JPCB2009,Leon-GutierrezEtal:TA2009,DawsonEtal:JPCL2011,%
WhitakerEtal:JPCB2013}. In simulations, one can determine directly the
average potential energy per particle and use that to assess stability~\cite{%
SinghEtal:NM2013,LyubimovEtal:JCP2013,LinEtal:JCP2014}. The enthalpy of a PVD
glass can be
compared to the measured or extrapolated enthalpy of an ordinary glass
that has been aged over a long period of time, thereby providing an indirect
means of assessing the age and stability of a material. Such a measure of
stability, however, is purely thermodynamic. It is therefore of interest to
determine whether dynamic measures of stability, such as characteristic
relaxation times, provide the same view of PVD glasses that, up to now, has been
generated on the basis of largely thermodynamic quantities. Annealing
experiments have demonstrated that, upon heating, stable glasses take a much longer time
to reach the liquid state than ordinary glasses, thereby suggesting that they exhibit strongly reduced dynamics~%
\cite{ChenEtal:JCP2013,SepulvedaEtal:JCP2013,SepulvedaEtal:PRL2014}.
Furthermore, dielectric measurements reveal a strong suppression of the
$\beta$-relaxation in stable glasses~\cite{YuEtal:PRL2015}. Similar techniques, however,
have not been applied in simulations of vapor-deposited glasses.
Here, a connection between the inherent structure energy and the dynamics is
of particular interest as the energy, similar to the enthalpy, is easily
accessible, whereas long and demanding simulation runs are necessary to extract
dynamic properties.

More generally, in this work we address the issue of whether vapor deposited
glasses are dynamically equivalent to aged glasses. We examine
whether the dynamics in vapor deposited glasses are comparable to those of
glasses aged over
long periods of time. We ask if vapor deposited glasses are indeed closer to
the equilibrium state than ordinary glasses, or if they represent a ``hidden
amorphous state''~%
\cite{DawsonEtal:PNAS2009} that transforms back to an ordinary glass over time.

To approach these questions, we analyze the decay of the incoherent intermediate
scattering
function (ISF) in vapor deposited and ordinary glasses for three different glass
formers: a bead-spring polymer melt, a two-dimensional binary mixture, and a
three-dimensional binary mixture. For all three systems, we find that the
dynamics
are indeed strongly slowed in stable glasses. Furthermore, we confirm
that
none of the models considered here
displays any sign of a ``hidden amorphous state''. Instead, we find that
the inherent structure energy is a good indicator for the dynamics, and that
slowly cooled or aged ordinary glasses with the same inherent structure energy
as a vapor deposited glass display almost identical dynamics. This finding also holds
for vapor deposited and liquid cooled polymer films, which are structurally
different.

Our manuscript begins with a summary of the simulation techniques employed in
this work
(\sref{sec:methods}), followed a discussion of the corresponding
results (\sref{sec:results}). We conclude with general remarks pertaining to the
stability and anisotropy of vapor deposited glasses (\sref{sec:conc}).

\section{Methods}
\label{sec:methods}

The models and simulation protocols used to replicate
the vapor deposition process have been described in the literature
for the polymer system\cite{LinEtal:JCP2014}, the 3D Lennard-Jones binary
mixture (3dBM)\cite{SinghEtal:NM2013,%
LyubimovEtal:JCP2013}, and the 2D Lennard-Jones binary system
(2dBM)\cite{Dan2D}.
For completeness, only a brief summary is provided in what follows. The first
glass former we consider is a binary
mixture consisting of two types of particles, type $A$ and type $B$, in a ratio
of $80/20$ for the 3dBM system and $65/35$ for the 2dBM system~%
\cite{BruningEtal:JPCM2009}.
The particles interact via a Lennard-Jones (LJ) potential truncated at
$r_{\text{trunc}} = 2.4$ and extrapolated to smoothly decay to zero at
$r_{\text{c}} = 2.5$. In the following, all values are reported in LJ units. To
this end, we set $\LJeps_{AA} = 1$, $\LJsigma_{AA} = 1$, the mass $m = 1$, and
$k_{\text{B}} = 1$. In these units, the relevant interaction
parameters are $\LJeps_{AB} = 1.5$, $\LJsigma_{AB} = 0.8$, $\LJeps_{BB} = 0.5$,
and $\LJsigma_{BB} = 0.88$.

As an alternative glass former, we study a simple bead-spring polymer model. The
polymer system consists only of type-$A$ particles, connected via bonds to
form chains of length $\chainlength = 4$ or $\chainlength = 10$, well
below the entanglement length~\cite{EveraersEtal:Science2004}. Bonded particles
(beads) are connected by a harmonic potential,
whose spring constant $K = 1000$ and equilibrium bond length $l_{0} =
0.97$ are chosen to prevent chain crossings and inhibit crystallization~%
\cite{GrestMurat:InBook1995}.
The substrate consists of a third type of atoms. It
interacts with the glass former via a LJ potential with the following
interaction parameters: $\LJeps_{AS} = \LJeps_{BS} = 0.1$ (3d BM), $\LJeps_{AS}
= \LJeps_{BS} = 1.0$ (2d BM, polymer), $\LJsigma_{AS} =
0.75$ (2d/3d BM), $\LJsigma_{AS} = 1.0$ (polymer), $\LJsigma_{BS} = 0.7$
(3d BM), $\LJsigma_{BS} = 0.75$ (2d BM), $\LJeps_{SS} = 0.1$, and $\LJsigma_{SS}
= 0.6$.

\begin{table}
    \begin{center}
        \setlength{\tabcolsep}{2pt}
    \begin{tabular}{cccccccc}
                & \multicolumn{3}{c}{Polymer $\chainlength = 4$} &
                  \multicolumn{4}{c}{Binary mixture (3d)}        \\
                  \hline
        $T$     & $0.3$   & $0.35$  & $0.4$             &
                  $0.275$ & $0.3$   & $0.325$ & $0.35$  \\
        $T/\Tg$ & $0.79$  & $0.92$  & $1.05$            &
                  $0.79$  & $0.86$  & $0.93$  & $1.00$  \\
                  \multicolumn{8}{c}{ } \\
                & \multicolumn{3}{c}{Polymer $\chainlength = 10$} &
                  \multicolumn{4}{c}{Binary mixture (2d)}         \\
                  \hline
        $T$     & $0.3$   & $0.35$  & $0.4$             &
                  $0.166$ & $0.182$ & $0.193$ & $0.221$ \\
        $T/\Tg$ & $0.73$  & $0.85$  & $0.98$            &
                  $0.79$  & $0.87$  & $0.92$  & $1.05$  \\
    \end{tabular}
    \end{center}
    \caption{Temperatures at which the ISF is determined}
    \label{tab:temp1}
\end{table}

To replicate the vapor deposition process, we first place $\Ns$ substrate
atoms randomly
in a thin layer at the bottom of the simulation box and
minimize the energy to remove overlap and to spread the atoms evenly across
the layer. Then, the substrate atoms are tethered to their current position
using a harmonic spring with spring constant $K = 1000$. Onto this substrate
we deposit the glass former. For the polymer glass and the 3dBM system, we
iterate the following steps: (1) We introduce
either a set $10$ particles
of
the binary mixture or one polymer chain
into the system and bring it into contact
with the surface of the film (or the substrate if no film has yet formed); (2)
We slowly cool the newly introduced particle(s) to the substrate temperature;
and
(3) We perform energy minimization using the FIRE algorithm~%
\cite{BitzekEtal:PRL2006}.
These steps are designed to improve the efficiency of the simulation by
assisting the newly deposited particle(s) in finding their most favorable,
nearby local energy minimum.
During deposition and the subsequent isothermal run, all particles
are coupled to an external
heat bath using the Nos\'{e}-Hoover thermostat~\cite{ShinodaEtal:PRB2004,%
FrenkelSmit:Book2002}.
This procedure differs
from the experimental situation in that local equilibrium is attained by
quenching the system to a nearby energy minimum through a steepest-descent
procedure.
By taking advantage of the reduced numerical
complexity afforded in two dimensions, for the 2dBM system we follow a more
realistic algorithm in which only
the substrate atoms are coupled to an external heat bath, and we refrain from
performing energy
minimizations. Instead, we allow newly deposited particles to reach the
equilibrium (substrate) temperature through ``natural'' energy dissipation
mechanisms.

\begin{figure}[tb]
    \centering
    \includegraphics[width=0.9\linewidth]{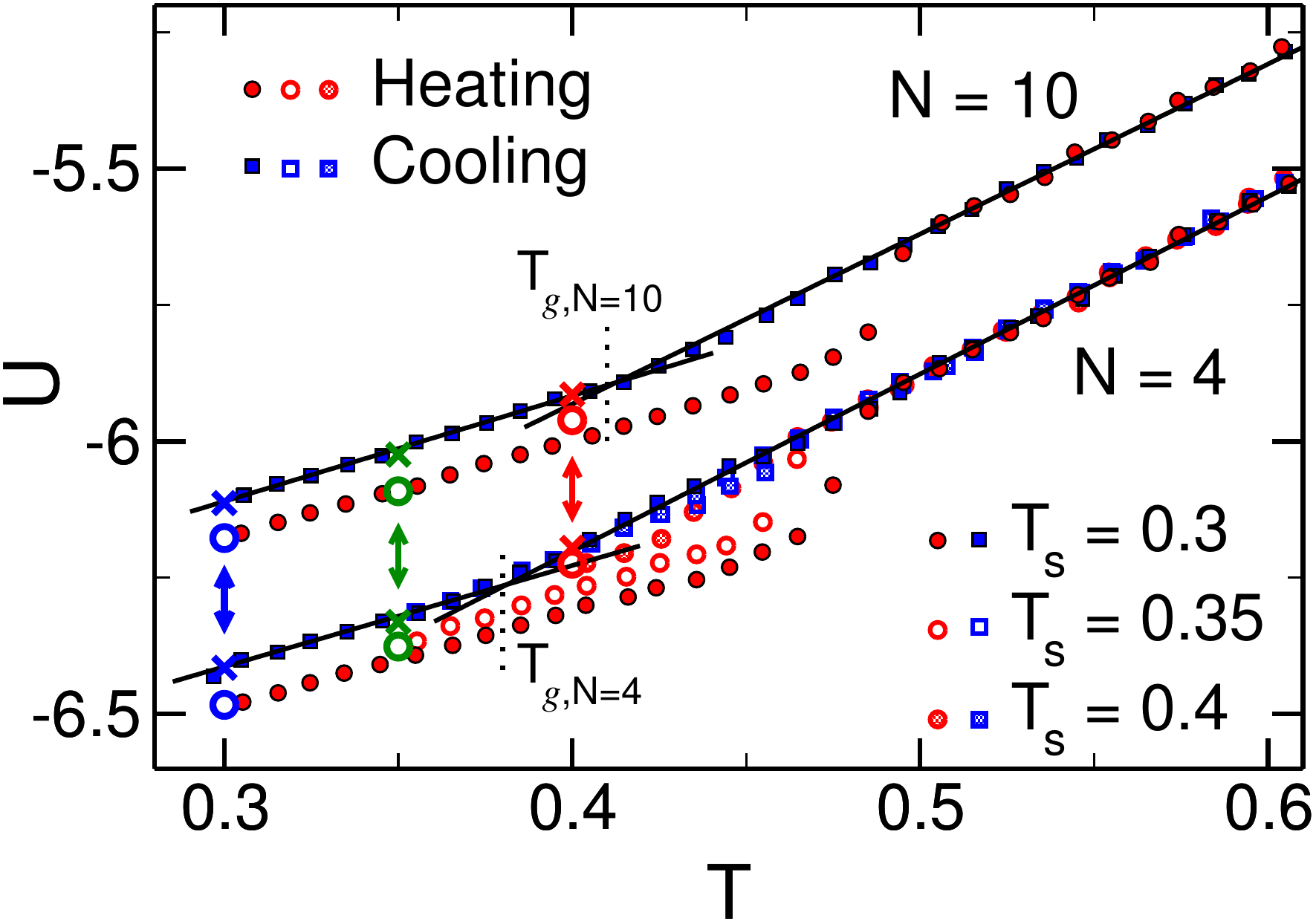}
    \caption{Average potential energy per particle, $U$, for the polymer model
        with chains of length $\chainlength = 4$ and $\chainlength = 10$ during
        a heating run from $\Ts$ to $T = 0.8$ (circles) and a subsequent cooling
        run back to $\Ts$ (squares). For both systems, the heating/cooling rate
        is $\coolrate = 10^{-5}$. For $\chainlength = 4$, the potential energy
        for systems deposited at $\Ts = 0.3$ (solid
        symbols), $\Ts = 0.35$ (open symbols), and $\Ts = 0.4$ (hatched symbols)
        are displayed. For $\chainlength = 10$ only the potential energy for the
        system deposited at $\Ts = 0.3$ is displayed. The potential energy has
        been averaged over all particles in the center region of the film and
        the results for $\chainlength = 10$ have been shifted by $0.2$ for
        clarity. The solid lines are fits to the linear regimes, the vertical
        lines indicate the values of $\Tg$ as defined in the main text. The arrows
        indicate the temperatures at which the ISF is calculated and the circles
        and crosses mark the potential energy at the start of the calculation for
        vapor deposited and ordinary glasses, respectively.}
    \label{fig:energy}
\end{figure}

After the deposition is complete, all films are relaxed for $500$ LJ-time
units before collecting data. Ordinary glass films are created by
heating vapor deposited films well above the glass transition temperature and
slowly cooling them to the desired temperature at a specified cooling rate.

In order to study the microscopic dynamics, we deposited films at various
substrate temperatures in the range from $0.73\Tg$ to $1.05\Tg$ (see
\tref{tab:temp1}). At
each temperature, we perform isothermal runs on
four independent configurations for each system.

\section{Results and discussion}
\label{sec:results}

We begin our discussion with an analysis of our results for the polymer model.
In order to compare vapor
deposited glasses to ordinary glasses, vapor deposited
materials are heated to $T = 0.8$, well above the glass transition temperature,
and cooled
back down to the desired temperature at a rate
of $\coolrate = 10^{-5}$. During this heating/cooling run, we calculate the
average potential energy in the center region of the film%
\cite{LinEtal:JCP2014}. Our aim is to
reproduce standard differential scanning calorimetry procedures, albeit at much
higher cooling rates than those typically used in experiments.
Our results for the energy, shown in \fref{fig:energy}, bear several of the
characteristics that have been experimentally
observed in stable glasses: A lower potential energy, and an
onset temperature $\Ton$ that is much higher than the glass transition
temperature $\Tg$ for the same cooling/heating rate.
These findings have been reported elsewhere, and strongly indicate that
vapor-deposited materials lie deeper in
the potential energy landscape\cite{LinEtal:JCP2014}. What has perhaps not been
firmly established
before, however, is whether a low potential energy does indeed translate
into slower structural relaxation in vapor deposited materials.

\begin{figure}[tb]
    \centering
    \includegraphics[width=0.9\linewidth]{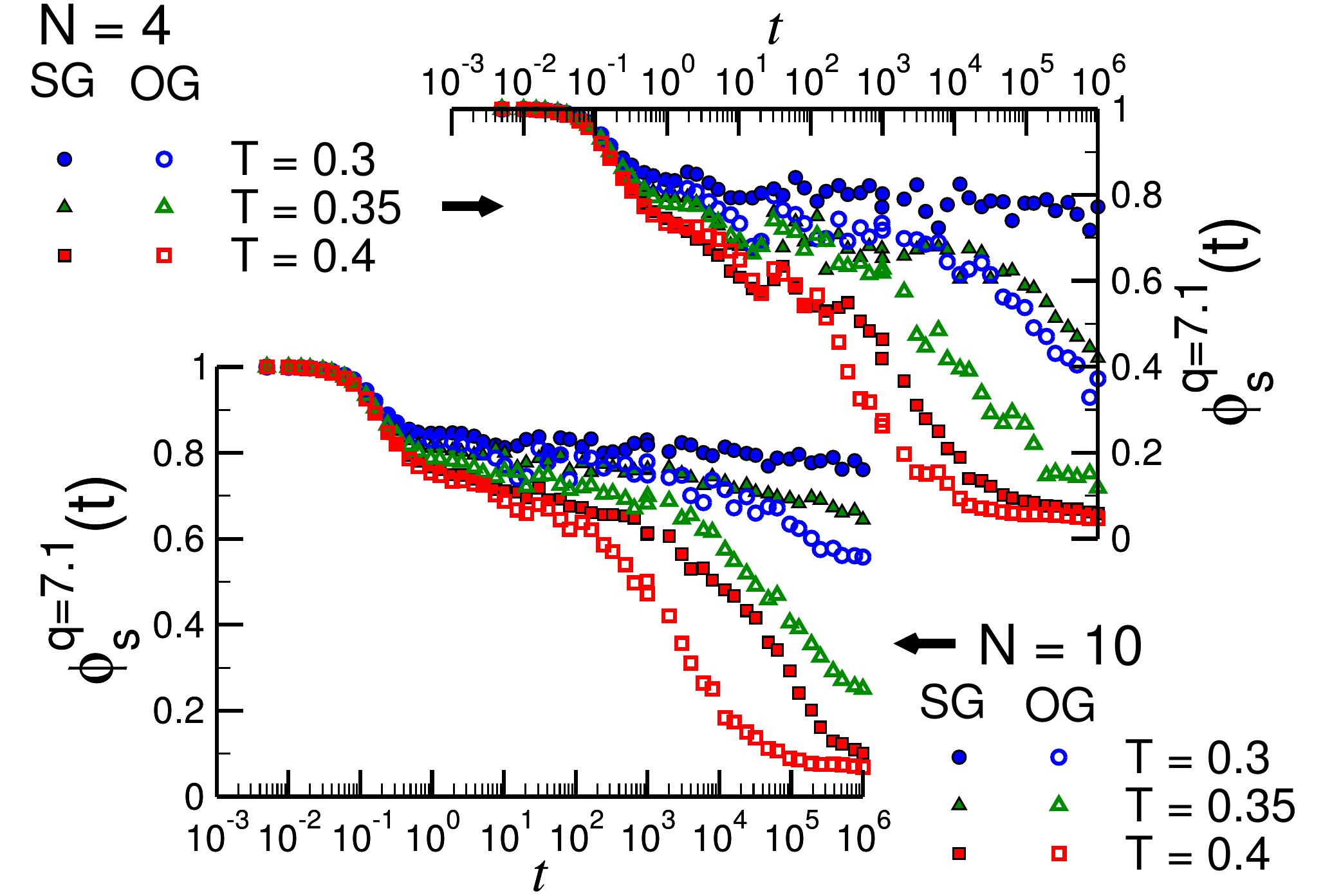}
    \caption{ISF $\isf(t)$ for chains of length $\chainlength = 4$ (top panel)
        and $\chainlength = 10$ (bottom panel). The solid symbols represent the
        results for the stable glass (SG), the open symbols those for the
        ordinary glass (OG). The ISF is determined at temperatures $T = 0.3$
        ($\circ$), $0.35$ ($\vartriangle$), and $0.4$ ($\square$).}
    \label{fig:isf}
\end{figure}

Our analysis of dynamics is performed by relying on the glass transition
temperature as a reference. In this work, we use the simulated potential energy
curves to define the relevant simulated glass transition temperatures $\Tg$ by
identifying two linear regimes in the cooling run, one corresponding to the
equilibrium
supercooled liquid and another to the glass regime, respectively. We then
fit the data in the linear regimes (solid lines in \fref{fig:energy}) and
calculate the intersection point. For the polymer systems considered here, we
find $\Tg = 0.38$ for $N = 4$
and $\Tg = 0.41$ for $N = 10$. For the binary systems, we find $\Tg = 0.35$ in
three dimensions\cite{LyubimovEtal:JCP2013} and $\Tg = 0.21$ in the
two-dimensional model\cite{Dan2D}.

The incoherent intermediate
scattering function (ISF) is given by
\begin{equation}
    \isf(t,\tage) = \frac{1}{N}\sum\limits_{k=1}^{N}
                        \langle
                            \exp\left[
                                i\vec{q}\left(
                                    r_{k}(\tage + t) - r_{k}(\tage)
                                \right)
                            \right]
                        \rangle \;,
    \label{eq:isf}
\end{equation}
where $\tage$ is the aging time, i.e.\ the time between the start of the
trajectory and the start of the measurement, and the average
$\langle\cdot\rangle$ is taken over $200$ $\vec{q}$-vectors in random
directions. Here, $\left|\vec{q}\right|$ corresponds to the first peak in the
static structure factor, which is at $\left|\vec{q}\right| = 0.71$ for the
polymer melt, $0.721$ for the 3dBM system, and $0.59$ for the 2dBM system. For
simplicity, we use $\isf(t, \tage = 0) \equiv \isf(t)$.

Our results for the polymer system are shown in \fref{fig:isf}. First, we note
that our
basic hypothesis holds true: the decay of the ISF for vapor-deposited materials
is slower than for their ordinary counterparts, indicating that the dynamics of
simulated vapor-deposited glasses are significantly
slower. This observation holds for all
temperatures and both chain lengths. Furthermore, we note
that the plateau value is slightly larger for the vapor deposited glass compared
to the respective liquid-cooled one. This observation is consistent with the
increased density observed in vapor-deposited systems. It is important to
emphasize that, at least for the $N=10$ polymeric system, the structure of vapor
deposited glasses is highly anisotropic and considerably different from that of
the corresponding ordinary material~\cite{LinEtal:JCP2014}. As such, it is
particularly important to stress that for these materials, a lower potential
energy continues to correspond to slower dynamics, even if the material adopts
different molecular packing arrangements. This finding is also relevant in view
of recent findings, which indicate that vapor-deposition enables control of
molecular anisotropy in stable glasses~\cite{DawsonEtal:JCP2012,%
BhattacharyaSadtchenko:JCP2014,IshiiNakayama:PCCP2014}.

\begin{figure}[tb]
    \centering
    \includegraphics[width=0.9\linewidth]{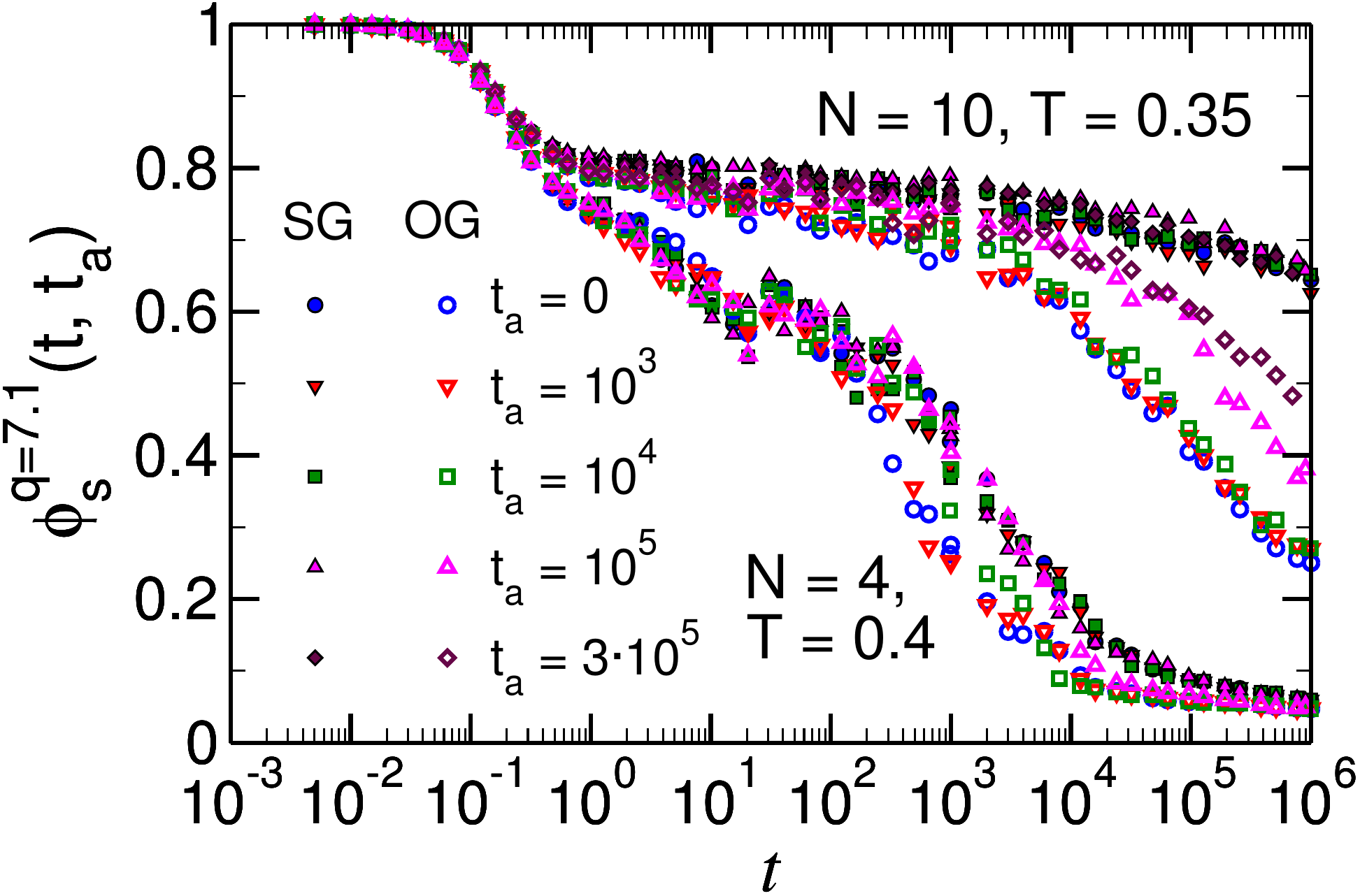}
    \caption{ISF $\isf(t,\tage)$ for chains of length $N = 4$ at $T = 0.4$ and
        $N = 10$ at $T = 0.35$. The ISFs are calculated both for the stable
        glass (SG, solid symbols) and the ordinary glass (OG, open symbols) at
        various aging times $\tage = 0$ ($\circ$), $10^{3}$ ($\triangledown$),
        $10^{4}$ ($\square$), and $10^{5}$ ($\vartriangle$), as well as $\tage =
        3\cdot 10^{5}$ ($\lozenge$) for chain of length $\chainlength = 10$.}
    \label{fig:isf_aging}
\end{figure}

We further note that dynamics in the vapor deposited glass is significantly
slower than in the ordinary material, even
when deposited above $\Tg$, as can be seen in \fref{fig:isf} for chains of
length $N = 4$ (deposited at $T = 0.4$). Given the fact that the potential
energy at $T = 0.4$ is very close to the equilibrium supercooled-liquid line, the
question arises as to which of the two systems corresponds to the equilibrium
state. To address this issue, we calculate the ISF for various aging
times. \Fref{fig:isf_aging} shows results for $\chainlength = 4$ at
$T = 0.4$ and $\chainlength = 10$ at $T = 0.35$. First, we note that at $T =
0.4$, the vapor deposited glass is in equilibrium, whereas for the fast cooling
rates considered here, the system cooled
from the liquid state retains some memory from its process of formation. We
attribute this behavior to the polymeric nature of the molecules: while the
system is able to relax on the scale
of the individual chain segments at $T = 0.4$, it is not able to do so on the
length scale of the entire chains. Our observation is further supported by
the fact that
the short chains deposited at $T = 0.4$ show an onset temperature slightly larger
than the glass transition temperature (see \fref{fig:energy}). For the longer chains
$\chainlength = 10$, below $\Tg$ equilibrium can no longer be attained
by aging on the time scales considered here. Instead, we observe the typical
aging behavior in which
the plateau value increases and the decay from the plateau
shifts to later times. In contrast, for the vapor deposited material, we find no
evidence of aging on the time scales accessible to our simulations: the ISF data
remain unchanged
for all aging times studied in this work. Given that the potential
energy of the stable glass lies on the extrapolated supercooled liquid line, it
is plausible to assume that the system has reached its equilibrium potential
energy. This
observation
further underscores that vapor deposition provides a surprisingly effective
experimental and simulation technique for preparation of glasses whose
properties are difficult to attain
via traditional liquid-cooling processes.

\begin{figure}
    \centering
    \includegraphics[width=0.9\linewidth]{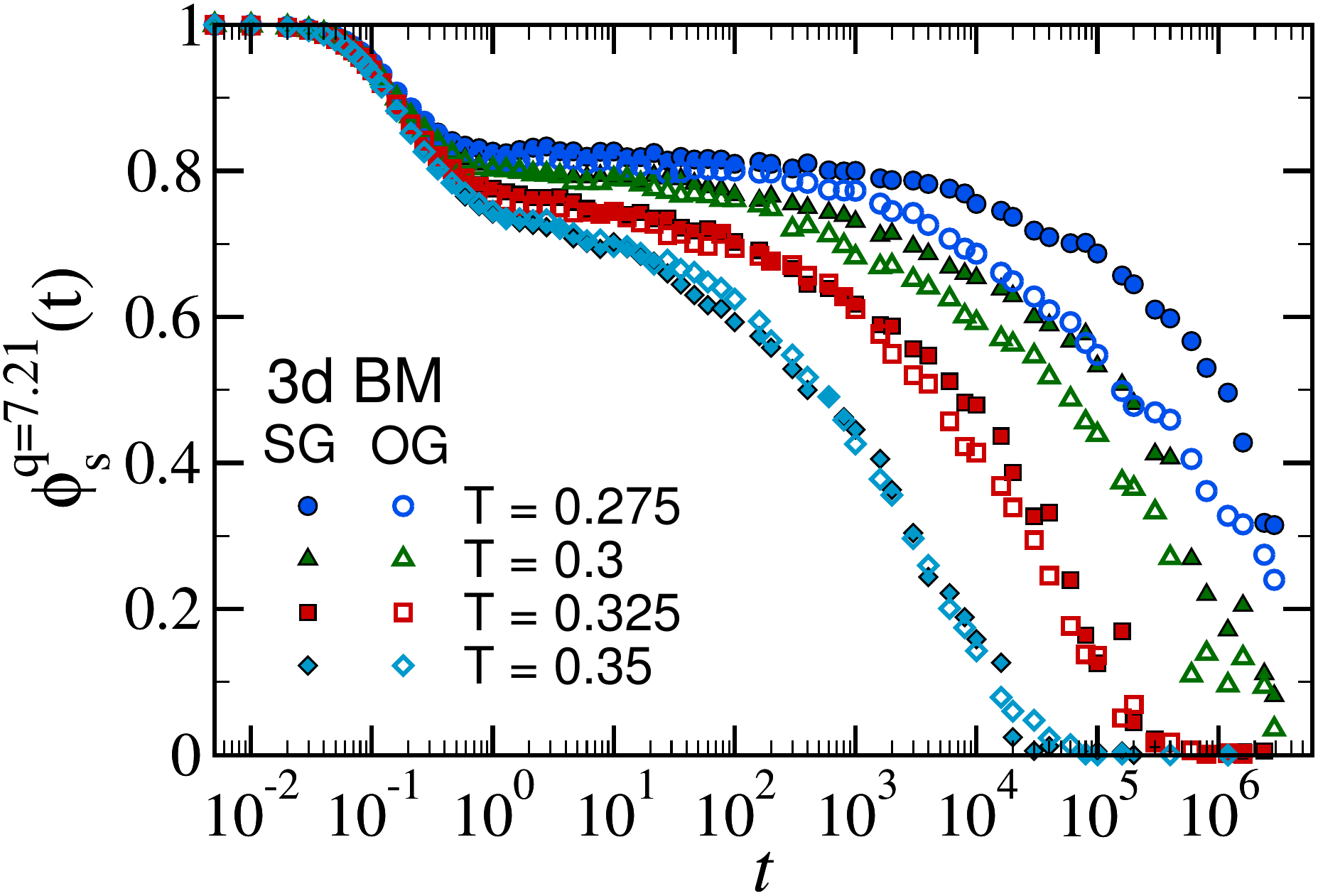}
    \caption{ISF $\isf(t)$ for 3d binary mixture stable glass (SG, full symbols)
        and ordinary glass (OG, open symbols) films at four different
        temperatures: $T=0.275$ ($\circ$), $0.3$ ($\vartriangle$), $0.325$
        ($\square$), and $0.350$ ($\lozenge$).}
    \label{fig:isf_bmlj_3d}
\end{figure}

We now consider whether the 3D and 2D binary
mixtures prepared by vapor deposition also exhibit slower dynamics. The ISFs for
the 3-dimensional system are
shown in \fref{fig:isf_bmlj_3d}. First,
we note that the general behavior is similar to that found for the polymer
system, albeit the
difference between ordinary and stable glasses is less pronounced. As with the
polymers, the difference in dynamics
between the vapor deposited and the ordinary glass increases with lower
deposition temperatures.

\begin{figure}
    \centering
    \includegraphics[width=0.9\linewidth]{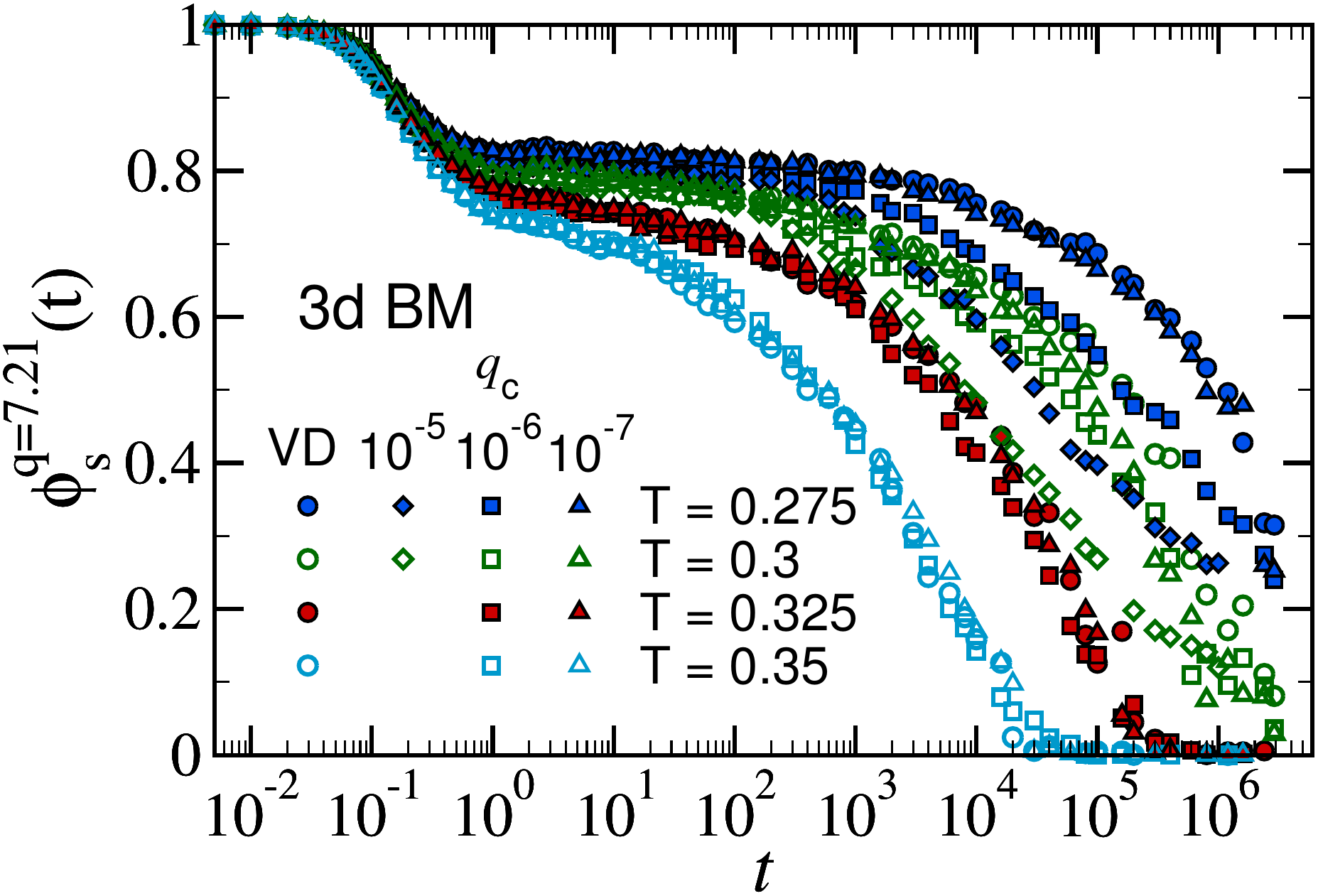}
    \caption{ISF $\isf(t)$ for 3d binary mixture of a vapor deposited glass (VD,
        $\circ$) and ordinary glasses formed using three different cooling rates
        $\coolrate = 10^{-5}$ ($\vartriangle$, shown only for the two lowest
        temperatures), $10^{-6}$ ($\square$), and $10^{-7}$ ($\lozenge$). The
        films were deposited at or cooled to four different temperatures:
        $T=0.275$, $0.3$, $0.325$, and $0.350$.}
    \label{fig:cooling_rate}
\end{figure}

For the polymer systems we examined the effect of aging on the material. For the
LJ mixtures, we examine the effect of cooling rate. Specifically, we consider
the following three
rates $\coolrate = 10^{-5}$, $10^{-6}$,
and $10^{-7}$. The corresponding ISFs are displayed in
\fref{fig:cooling_rate}. As expected, the dynamics are markedly slower for
systems cooled by slower
cooling rates. For the slowest cooling rate considered here, the dynamics almost
overlap with those of the
as-deposited films. We note, however, that the liquid-cooling procedure is much
more computationally demanding than vapor deposition.

\begin{table}
    \begin{center}
    \begin{tabular}{ccccc}
        $T$                  & $0.166$  & $0.182$  & $0.193$  & $0.221$ \\
        \hline
        \vspace{3pt}
        $U^{\text{IS}}_{\text{VD}}$ & $-3.719$ & $-3.710$ & $-3.704$ & $-3.686$ \\
        \vspace{3pt}
        $U^{\text{IS}}_{\text{c}2}$ & $-3.719$ & $-3.716$ & $-3.701$ & $-3.688$ \\
        \vspace{3pt}
        $U^{\text{IS}}_{\text{c}1}$ & $-3.716$ & $-3.708$ & $-3.696$ & $-3.690$ \\
        \hline
    \end{tabular}
    \end{center}
    \caption{Inherent structure energy $U^{\text{IS}}$ for the vapor deposited
        film (VD) and two liquid-cooled films using the cooling rates
        $q_{\text{c}2} = 2\cdot 10^{-7}$ and $q_{\text{c}1} = 1.33\cdot
        10^{-6}$. The cooling rates are chosen such that (1) the cooling time is
        identical to the deposition time or (2) the inherent structure energy of
        vapor deposited and liquid cooled films are approximately equal.}
    \label{tab:temp}
\end{table}

These results do, in fact, raise a more fundamental question: What is a fair
comparison between vapor deposited and liquid cooled films? This question is
particularly important for numerical simulations, where typical cooling rates
are orders of magnitude faster than the 1 $K$/min typically employed in
experiments. Two
measures can be considered: One can (1) compare cooling and deposition protocols
that are run over the same amount of time, measured in LJ time units, or, (2),
which processes lead to
the same inherent structure energy, i.e.\ which glasses
are comparable in terms of the potential energy landscape.
The lowest inherent structure energies that can be reached by vapor deposition
are, in general, not attainable by liquid cooling. To reach the lowest inherent
structure energy reported in ~\rref{Dan2D} by liquid cooling,
a rate on the order of $10^{-13}$ would be necessary, requiring on the order
of thousands of years of simulation time on a typical CPU. For the purposes of this
manuscript, we have thus deliberately chosen suboptimal parameters for the vapor deposition
process, leading to rather ''poor'' glasses in comparison. Only these suboptimal parameters allow
the liquid cooled system to reach in reasonable timescales the same inherent structure energy that is achieved by vapor deposition.
The inherent structure energies for the three different
protocols are listed in \tref{tab:temp}. We note that the inherent structure
energies are lower for the vapor deposited films compared to the liquid cooled
ones formed during the same amount of time, except for the highest temperature
which is above $\Tg$. This demonstrates that vapor deposited glass lie deep in
the potential energy landscape.

\begin{figure}
    \centering
    \includegraphics[width=0.9\linewidth]{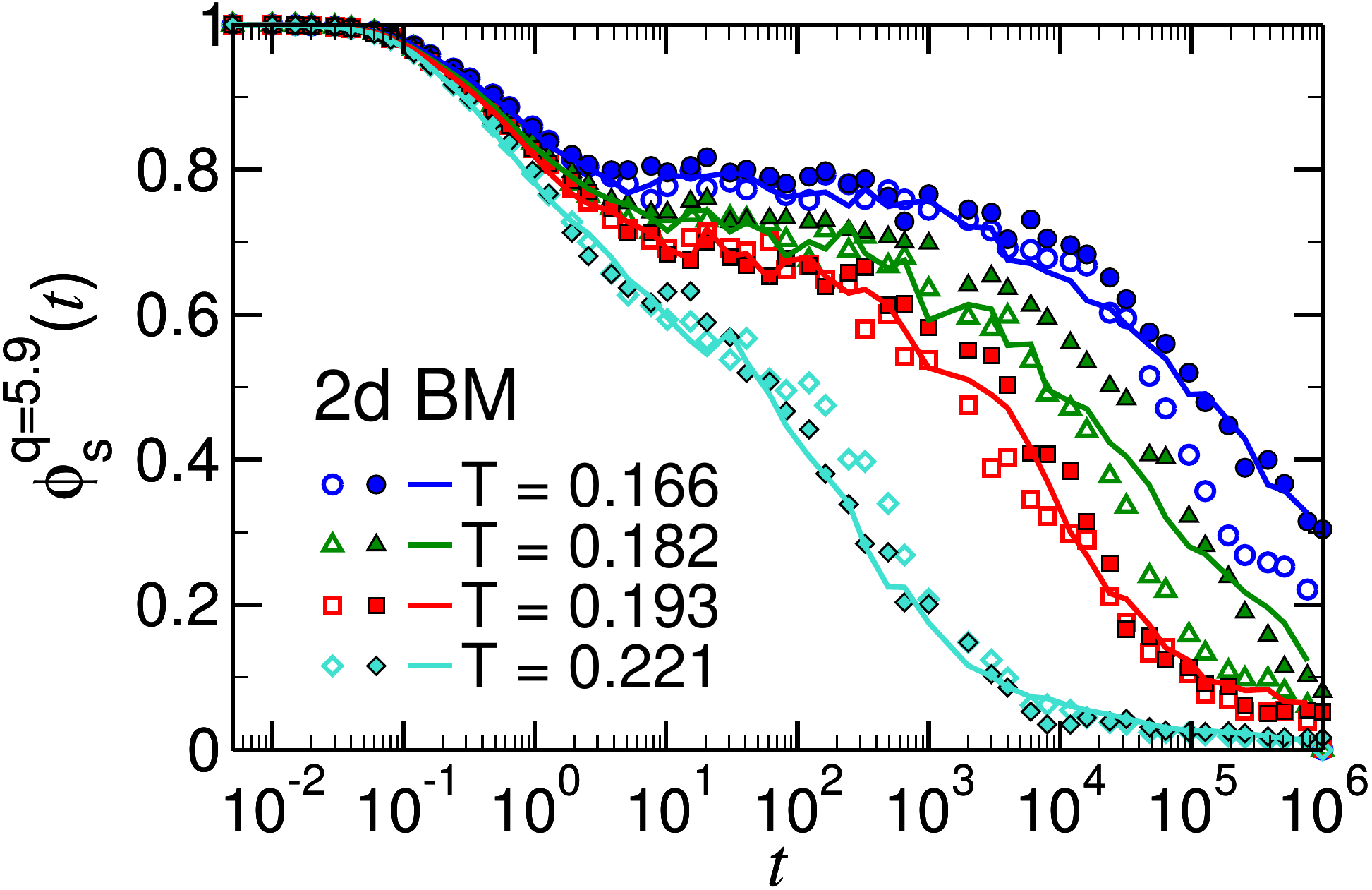}
    \caption{ISF $\isf(t)$ for a 2d binary mixture of a vapor deposited glass
        (lines) and two ordinary glasses formed using two different
        cooling rates $\coolrate$. The cooling rates were chosen such that the
        ordinary glass forms in the same amount of time as the vapor deposited
        one (open symbols, $\coolrate = 1.33\cdot 10^{-6}$) or such that both
        films obtain the same inherent structure energy (solid symbols,
        $\coolrate = 2\cdot 10^{-7}$). The glasses were deposited at or cooled
        to four temperatures $T=0.166$ ($\circ$), $0.182$ ($\vartriangle$),
        $0.325$ ($\square$), and $0.350$ ($\lozenge$).}
    \label{fig:isf_2d_bm}
\end{figure}

The ISFs for the three different formation protocols are displayed in
\fref{fig:isf_2d_bm}. Two conclusions can be drawn from the figure.
First, at lower temperatures, the ISF decays faster for the liquid cooled
system formed over the same amount of time as the vapor deposited film.
This further substantiates our premise that vapor deposition provides an
efficient
means by which to form glasses that reside deep in the potential energy
landscape. Second,
glasses that reach the same inherent structure energy via slow liquid cooling or
vapor deposition exhibit almost identical dynamics. In fact, the ISFs
mostly overlap with $T = 0.182$ being a notable outlier. This is,
however, not surprising given the fact that the inherent structure energy is
lower for the slowly cooled system at this temperature (see \tref{tab:temp}).
Thus, \fref{fig:isf_2d_bm} suggests that systems with the same inherent
structure energy display identical dynamics as measured by the incoherent
scattering function. This is far from obvious, considering the different formation
protocols, yet it is not implausible given that both systems are
structurally similar.

\begin{figure}[t]
    \centering
    \includegraphics[width=0.9\linewidth]{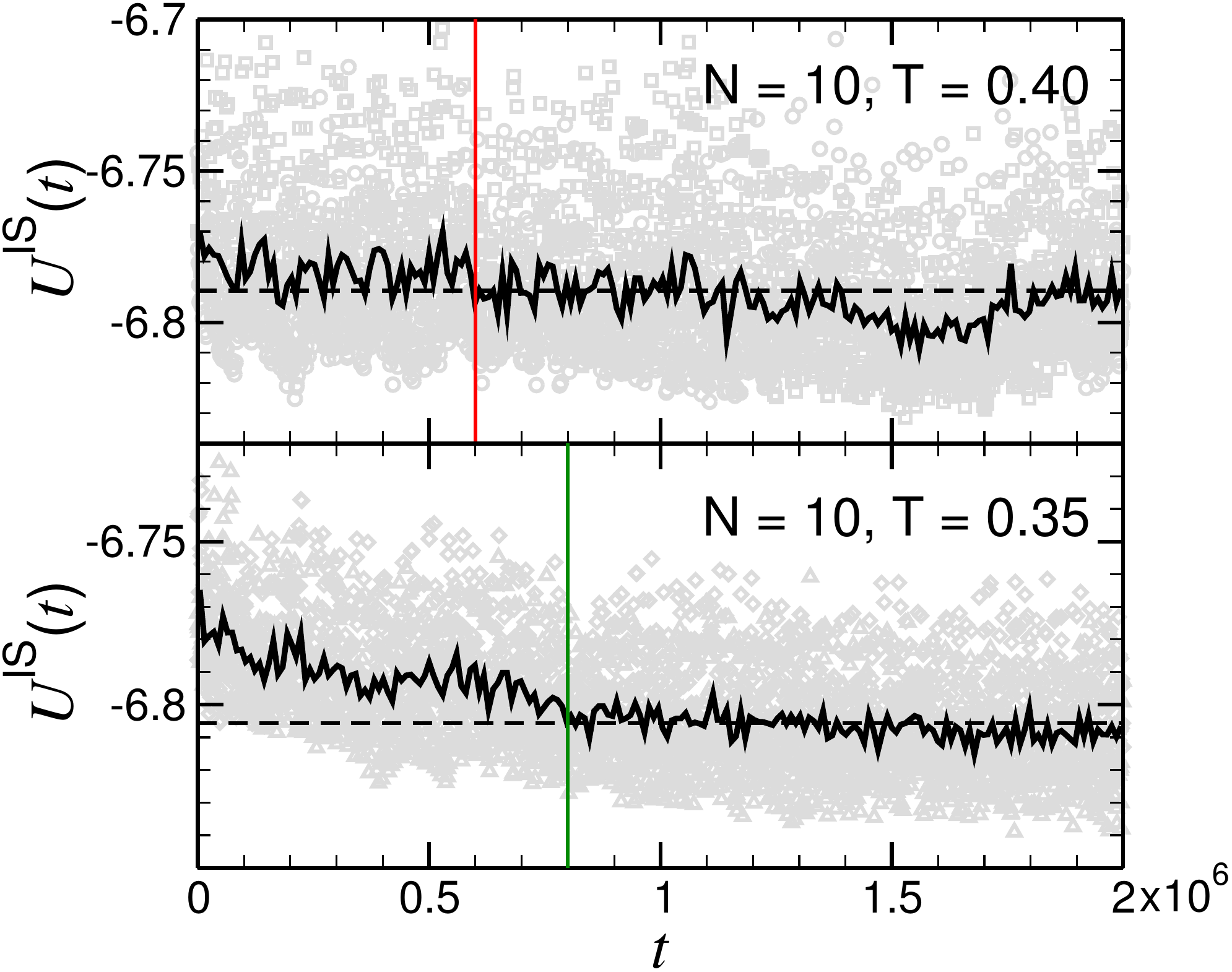}
    \caption{Inherent structure energy $U^{\text{IS}}(t)$ for the ordinary glass
    cooled to $T = 0.40$ (upper panel) and $0.35$ (lower panel). The solid line
    displays $U^{\text{IS}}$ averaged over $20$ trajectory points, $10$ each from
    two independent configurations. $U^{\text{IS}}$ for the individual trajectory
    points is displayed as grey symbols. The horizontal dashed line indicates the
    average inherent structure energy of the just-deposited stable glass deposited
    at $\Ts = 0.40$ (upper panel) and $0.35$ (lower panel). The vertical lines
    indicate the times at which the ordinary glass reaches approximately the
    inherent structure energy of the vapor deposited glass directly after
    deposition.}
    \label{fig:is_energy}
\end{figure}

To test whether this finding holds also for systems that are structurally
different, we turn
again to the polymer system with chains of length $N = 10$. During deposition,
these polymers stretch out on the surface and retain their alignment as the film
grows~\cite{LinEtal:JCP2014}. Thus, vapor deposited polymer films display strong
anisotropy, with the end-to-end vectors of the polymers closely aligned parallel
to the film surface. Liquid-cooled polymer films, on the other hand, are fully
isotropic. To determine the inherent structure energy $U^{\text{IS}}$, we
performed an isothermal run and minimized the energy at regular intervals
$\Delta t$ using the FIRE~\cite{BitzekEtal:PRL2006} algorithm. We determined the
average $U^{\text{IS}}$ for the vapor deposited glass within the first 1000 LJ
time units ($\Delta t = 100$). Next, we determined $U^{\text{IS}}(t)$ for the
ordinary glass over a long trajectory ($\Delta t = 1000$) and estimated the time
at which the inherent structure energy reached the same value as the
just-deposited film. The results are displayed in \fref{fig:is_energy}. At $T =
0.40$, i.e.\ above the glass transition temperature, $U^{\text{IS}}(t)$ changes
only a little, whereas at $T = 0.35$ a clear trend is visible. We estimate that
after a time $\tage = 6\cdot 10^{5}$ for the system above $\Tg$ and $\tage =
8\cdot 10^{5}$ for the system below $\Tg$, the ordinary glass has the same
inherent structure energy as the vapor deposited glass. As for the 2d-KA
system, we have again deliberately chosen suboptimal parameters for the
vapor deposited glass at $T = 0.35$, using a deposition rate that is about one
order of magnitude faster than that for all other polymer systems considered here.
It is only with this very fast deposition rate that we arrive at an inherent structure
energy that is sufficiently high to be attainable by simple aging of an ordinary glass.

\begin{figure}[t]
    \centering
    \includegraphics[width=0.9\linewidth]{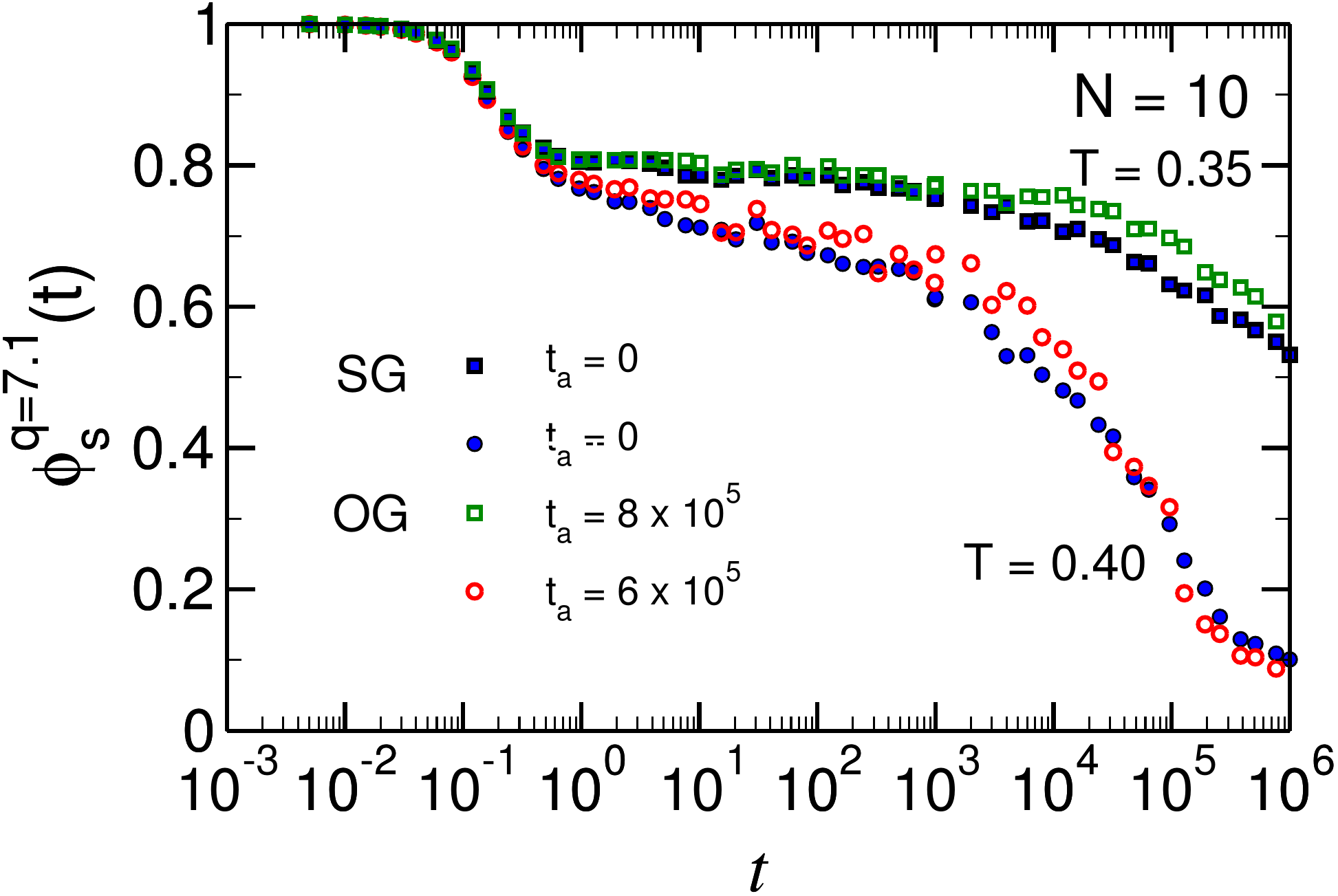}
    \caption{ISF $\isf(t)$ for the vapor deposited glass (SG, solid symbols) and
    the ordinary glass (OG, open symbols) deposited at or cooled to $T = 0.40$
    ($\circ$) and $0.35$ ($\square$). The ordinary glass was aged at $T = 0.40$
    and $0.35$ for $t_{a} = 5\cdot 10^{5}$ and $8\cdot 10^{5}$, respectively.
    The aging times were chosen such that both ordinary and stable glass have
    the same inherent structure energy (see \fref{fig:is_energy}).}
    \label{fig:isf_same_is}
\end{figure}

The results for the ISF are displayed in \fref{fig:isf_same_is}. We find that
the ISFs of the vapor deposited glass and the ordinary glass are essentially identical,
given the fact that they exhibit the same inherent structure energy. It
is important to emphasize that this does not imply that the systems are in the same
state. In fact, given the structural differences, we expect them to exhibit different
mechanic and thermodynamic properties, and it would not be surprising if they
displayed a different aging behavior. However, our results suggest that the
inherent structure energy is a good indicator to gauge the local dynamics. Thus,
we infer that it is also a good measure of the stability of the glass. Furthermore,
this finding can also be interpreted in terms of the potential energy landscape.
The similar dynamics of the systems, despite their different structures, suggests
that they are not only in a potential that has the same depth (i.e. the same
inherent structure energy), but also that the potential energy barriers
surrounding them have similar heights, leading
to similar dynamics. This, surprisingly, appears to hold
even though pronounced structural differences indicate that the systems are in very
different areas of the potential energy landscape. In other words, one could also argue
that in deep energy minima, a different structure does not necessarily imply that the potential energy
landscape exhibits a considerably different shape.

\section{Conclusions}
\label{sec:conc}

In this paper, we have demonstrated that the increased thermodynamic stability
of simulated vapor deposited glasses is also manifest in the microscopic
dynamics, as measured
by the incoherent intermediate scattering function. These dynamics resemble
closely those observed in
ordinary glasses aged for long periods of time or liquid-cooled materials
prepared at slow cooling rates. This
may not come as a surprise. Indeed, vapor deposited glasses have been frequently
compared to glasses aged for long times~\cite{KearnsEtal:JCP2007,%
KearnsEtal:JPCB2008}. Note, however, it is not immediately obvious that vapor
deposited
and well-aged glasses should exhibit identical
dynamics. Indeed, a contrasting hypothesis would be that a vapor deposition
process puts the glass in a ``stable state'' that is not accessible via liquid
cooling,
from which the system would transform back into an ordinary glass. In fact,
experiments suggest that some organic molecules are indeed deposited in a
``hidden amorphous state'' not accessible via liquid cooling~\cite{%
DawsonEtal:PNAS2009}. However, the results presented here show no evidence for
such states, neither for a bead-spring polymer model, where chains remain in
an anisotropic state, nor for a binary mixture which is, by construction, fully
isotropic. We have compared the ISFs of vapor-deposited glasses to those of ordinary
glasses either aged for various times or cooled at different cooling rates.
We find that the dynamics of ordinary glasses gradually approach those of a vapor deposited glass for
longer aging times and slower cooling rates. Here we note that
vapor deposition represents a very different and efficient process for preparation of well-equilibrated
glasses, both in experiments and in numerical simulations. Finally, we also
compared vapor deposited glasses and ordinary glasses where either the cooling
rate or the aging time were adjusted such that their final states would
have the same inherent structure energy. We found that these systems display the
same dynamics, even when comparing a highly anisotropic vapor deposited polymer film
to a structurally different, isotropic, liquid-cooled one. This finding supports the
assumption that the inherent structure energy is a good measure of the
stability of a glass.

\subsection*{Acknowledgements}
\label{sec:acknowledgements}

We gratefully acknowledge insightful discussions with Mark Ediger. JH acknowledges the financial support from the DFG research fellowship
program, grant No.~HE~7429/1. This work is supported by the National Science Foundation under a DMREF grant.

\bibliography{references}

\begin{thebibliography}{10}

\bibitem{BerthierBiroli:RMP2011}
L.~Berthier and G.~Biroli.
\newblock Theoretical perspective on the glass transition and amorphous
  materials.
\newblock {\em Rev. Mod. Phys.}, 83:587, 2011.

\bibitem{EdigerHarrowell:JCP2012}
M.~D. Ediger and P.~Harrowell.
\newblock Perspective: Supercooled liquids and glasses.
\newblock {\em J. Chem. Phys.}, 137:080901, 2012.

\bibitem{BiroliGarrahan:JCP2013}
G.~Biroli and J.~P. Garrahan.
\newblock Perspective: The glass transition.
\newblock {\em J. Chem. Phys.}, 138:12A301, 2013.

\bibitem{Donth:Book2001}
E.-J. Donth.
\newblock {\em The Glass Transition}.
\newblock Springer-Verlag, 2001.

\bibitem{BinderKob:Book2011}
K.~Binder and W.~Kob.
\newblock {\em Glassy Materials and Disordered Solids}.
\newblock World Scientific Publishing, 2011.

\bibitem{Struik:Book1978}
L.~C.~E. Struik.
\newblock {\em Physical aging in amorphous polymers and other materials}.
\newblock Elsevier, 1978.

\bibitem{SwallenEtal:Science2007}
S.~F. Swallen, K.~L. Kearns, M.~K. Mapes, Y.~S. Kim, R.~J. McMahon, M.~D.
  Ediger, T.~Wu, L.~Yu, and S.~Satija.
\newblock Organic glasses with exceptional thermodynamic and kinetic stability.
\newblock {\em Science}, 315:353, 2007.

\bibitem{IshiiNakayama:PCCP2014}
K.~Ishii and H.~Nakayama.
\newblock Structural relaxation of vapor-deposited molecular glasses and
  supercooled liquids.
\newblock {\em Phys. Chem. Chem. Phys.}, 16:12073, 2014.

\bibitem{DalalEdiger:JPCL2012}
S.~S. Dalal and M.~D. Ediger.
\newblock Molecular orientation in stable glasses of indomethacin.
\newblock {\em J. Phys. Chem. Lett.}, 3:1229, 2012.

\bibitem{KearnsEtal:JPCB2008}
K.~L. Kearns, S.~F. Swallen, M.~D. Ediger, T.~Wu, Y.~Sun, and L.~Yu.
\newblock Hiking down the energy landscape: Progress toward the kautzmann
  temperature via vapor deposition.
\newblock {\em J. Phys. Chem. B}, 112:4934, 2008.

\bibitem{KearnsEtal:JPCB2009}
K.~L. Kearns, S.~F. Swallen, M.~D. Ediger, Y.~Sun, and L.~Yu.
\newblock Calorimetric evidence for two distinct molecular packing arrangements
  in stable glasses of indomethacin.
\newblock {\em J. Phys. Chem. B}, 113:1579, 2009.

\bibitem{Leon-GutierrezEtal:TA2009}
E.~Le{\'o}n-Gutierrez, G.~Garcia, M.~T. Clavaguera-Mora, and
  J.~Rodr{\'i}guez-Viejo.
\newblock Glass transition in vapor deposited thin films of toluene.
\newblock {\em Thermochim. Acta}, 492:51, 2009.

\bibitem{DawsonEtal:JPCL2011}
K.~Dawson, L.~Zhu, L.~A. Kopff, R.~J. McMahon, L.~Yu, and M.~D. Ediger.
\newblock Highly stable vapor-deposited glasses of four tris-naphthylbenzene
  isomers.
\newblock {\em J. Phys. Chem. Lett.}, 2:2683, 2011.

\bibitem{SinghDePablo:JCP2011}
S.~Singh and J.~J. de~Pablo.
\newblock A molecular view of vapor deposited glasses.
\newblock {\em J. Chem. Phys.}, 134:194903, 2011.

\bibitem{SinghEtal:NM2013}
S.~Singh, M.~D. Ediger, and J.~J. de~Pablo.
\newblock Ultrastable glasses from in silico vapour deposition.
\newblock {\em Nature Materials}, 12:139, 2013.

\bibitem{LyubimovEtal:JCP2013}
I.~Lyubimov, M.~D. Ediger, and J.~J. de~Pablo.
\newblock Model vapor-deposited glasses: Growth front and composition effects.
\newblock {\em J. Chem. Phys.}, 139:144505, 2013.

\bibitem{LinEtal:JCP2014}
P.-H. Lin, I.~Lyubimov, L.~Yu, M.~D. Ediger, and J.~de~Pablo.
\newblock Molecular modeling of vapor-deposited polymer glasses.
\newblock {\em J. Chem. Phys.}, 140:204504, 2014.

\bibitem{LyubimovEtal:JCP2015}
I.~Lyubimov, L.~Antony, D.~M. Walters, D.~Rodney, M.~D. Ediger, and J.~J.
  de~Pablo.
\newblock Orientational anisotropy in simulated vapor-deposited molecular
  glasses.
\newblock {\em J. Chem. Phys.}, 143:094502, 2015.

\bibitem{KearnsEtal:JCP2007}
K.~L. Kearns, S.~F. Swallen, M.~D. Ediger, T.~Wu, and L.~Yu.
\newblock Influence of substrate temperature on the stability of glasses
  prepared by vapor deposition.
\newblock {\em J. Chem. Phys.}, 127:154702, 2007.

\bibitem{WhitakerEtal:JPCB2013}
K.~R. Whitaker, D.~J. Scifo, M.~D. Ediger, M.~Ahrenberg, and C.~Schick.
\newblock Highly stable glasses of cis-decalin and cis/trans-decalin mixtures.
\newblock {\em J. Phys. Chem. B}, 117:12724, 2013.

\bibitem{ChenEtal:JCP2013}
Z.~Chen, A.~Sepulveda, M.~D. Ediger, and R.~Richert.
\newblock Dynamics of glass-forming liquids. xvi. observation of ultrastable
  glass transformation via dielectric spectroscopy.
\newblock {\em J. Chem. Phys.}, 138:12A519, 2013.

\bibitem{SepulvedaEtal:JCP2013}
A.~Sepulveda, S.~F. Swallen, and M.~D. Ediger.
\newblock Manipulating the properties of stable organic glasses using kinetic
  facilitation.
\newblock {\em J. Chem. Phys.}, 138:12A517, 2013.

\bibitem{SepulvedaEtal:PRL2014}
A.~Sepulveda, M.~Tylinski, A.~Guiseppi-Elie, R.~Richert, and M.~D. Ediger.
\newblock Role of fragility in the formation of highly stable organic glasses.
\newblock {\em Phys. Rev. Lett.}, 113:045901, 2014.

\bibitem{YuEtal:PRL2015}
H.~B. Yu, M.~Tylinski, A.~Guiseppi-Elie, M.~D. Ediger, and R.~Richert.
\newblock Suppression of $\beta$-relaxation in vapor-deposited ultrastable
  glasses.
\newblock {\em Phys. Rev. Lett.}, 115:185501, 2015.

\bibitem{DawsonEtal:PNAS2009}
K.~J. Dawson, K.~L. Kearns, L.~Yu, W.~Steffen, and M.~D. Ediger.
\newblock Physical vapor deposition as a route to hidden amorphous states.
\newblock {\em Proc. Natl. Acad. Sci. USA}, 106:15165, 2009.

\bibitem{Dan2D}
D.~Reid.
\newblock Age and structure of a model vapor-deposited glass.
\newblock Submitted to Nature Communications, 2016.

\bibitem{BruningEtal:JPCM2009}
R.~Br{\"u}ning, D.~A. St-Onge, S.~Patterson, and W.~Kob.
\newblock Glass transitions in one-, two-, three-, and four-dimensional binary
  lennard-jones systems.
\newblock {\em J. Phys. Condens. Matter}, 21:035117, 2009.

\bibitem{EveraersEtal:Science2004}
R.~Everaers, S.~K. Sukumaran, G.~S. Grest, C.~Svaneborg, A.~Sivasubramanian,
  and K.~Kremer.
\newblock Rheology and microscopic topology of entangled polymeric liquids.
\newblock {\em Science}, 303:823, 2004.

\bibitem{GrestMurat:InBook1995}
G.~S. Grest and M.~Murat.
\newblock Computer simulations of tethered chains.
\newblock In {\em Monte Carlo and Molecular Dynamics Simulations in Polymer
  Science}. Oxford University Press, 1995.

\bibitem{BitzekEtal:PRL2006}
E.~Bitzek, Koskinen. P., F.~G{\"a}hler, M.~Moseler, and P.~Gumbsch.
\newblock Structural relaxation made simple.
\newblock {\em Phys. Rev. Lett.}, 97:170201, 2006.

\bibitem{ShinodaEtal:PRB2004}
W.~Shinoda, M.~Shiga, and M.~Mikami.
\newblock Rapid estimation of elastic constants by molecular dynamics
  simulation under constant stress.
\newblock {\em Phys. Rev. B}, 69:134103, 2004.

\bibitem{FrenkelSmit:Book2002}
D.~Frenkel and B.~Smit.
\newblock {\em Understanding Molecular Simulation}.
\newblock Academic Press, 2002.

\bibitem{DawsonEtal:JCP2012}
K.~Dawson, L.~A. Kopff, L.~Zhu, R.~J. McMahon, L.~Yu, R.~Richert, and M.~D.
  Ediger.
\newblock Molecular packing in highly stable glasses of vapor-deposited
  tris-naphthylbenzene isomers.
\newblock {\em J. Chem. Phys.}, 136:094505, 2012.

\bibitem{BhattacharyaSadtchenko:JCP2014}
D.~Bhattacharya and V.~Sadtchenko.
\newblock Enthalpy and high temperature relaxation kinetics of stable
  vapor-deposited glasses of toluene.
\newblock {\em J. Chem. Phys.}, 141:094502, 2014.

\end{thebibliography}
\bibliographystyle{unsrt}

\end{document}